\documentclass{aa}
\usepackage{amssymb}

\usepackage{psfig}

\begin{document}

   \thesaurus{01     % A&A Section 1: Letters
	 (09.01.1;  % ISM:abudances,
	  09.03.1;  % ISM:clouds,
	  09.19.1;  % ISM:structure,
	  08.23.3;  % Stars:winds,outflows 
	  06.19.2;  % Solar wind;
	  13.21.3)} % Ultraviolet:ISM 

\title{Heliospheric and \textit{%
Astrospheric} Hydrogen Absorption towards Sirius: No Need for Interstellar
Hot Gas \thanks{
Based on observations with the NASA/ESA Hubble Space
Telescope, obtained at the Space Telescope Science Institute which is
operated by the Association of Universities for Research in Astronomy Inc., 
under NASA Contract NASA5-26555}
}
\author{V. V. Izmodenov  \inst{1} \fnmsep \thanks{
Permanently at Moscow State University, Department of Aeromechanics and Gas
dynamics, Faculty of Mechanics and Mathematics, E-mail: izmod@ipmnet.ru }
\and R. Lallement \inst{1}
\and Y. G. Malama \inst{2}
}

\offprints{R. Lallement}

\institute{Service d'Aeronomie, CNRS,
91371, Verrieres le Buisson, France \\
E-mail: Vlad.Izmodenov@aerov.jussieu.fr, Rosine.Lallement@aerov.jussieu.fr
\and
Institute for Problems in Mechanics, Russian Academy of Science, Moscow,
Russia
}

\authorrunning{Izmodenov \& al.}
\titlerunning{Hydrogen Absorption towards Sirius}
\date{Received ; accepted }
\maketitle

\begin{abstract}
We use an updated self-consistent kinetic/ gasdynamic model of the solar
wind/interstellar flow interaction to compute the Lyman-$\alpha$ absorption
by hydrogen atoms of both interstellar and solar origin generated in the
heliospheric interface.  We apply this model to the direction of the star
Sirius.  We show that the neutralized, compressed solar wind from the
heliosheath explains the extra absorption at positive redshifts observed in 
the Lyman-$\alpha$ line with the HST/GHRS.  This removes the need for the
interstellar hot gas previously proposed to explain this extra absorption.
We also show that extra absorption in the blue wing can possibly be
explained by H atoms formed in an 
\textit{%
astrosphere} around Sirius, providing the stellar wind is at least as
massive and fast as the solar wind.
\end{abstract}

%________________________________________________________________

\section{Introduction}

There is a long-standing debate on the properties of the hot ( $T \approx
10^6 $ K) gas which fills the so-called Local Bubble, the soft X-ray
emitting region around the Sun (see Breitschwerdt et al.\ 1997 for updated
information). The cold and warm diffuse clouds embedded in the hot gas,
like the group of cloudlets in the solar vicinity, are supposed to be
surrounded by a shell of semi-hot gas at an intermediate temperature ($T
\approx 10^5$ K), and one way to detect this gas is through the neutral
hydrogen absorption at Ly-$\alpha$.  It is worthwhile here to recall that a
small quantity of hot gas ($T\approx 10^5$ K) can produce an absorption
equivalent width in Ly-$\alpha$ comparable to that of a 3
or 4 orders of magnitude larger quantity of warm gas ($T\approx 10^4$ K).

Indeed, spectra obtained with the Goddard High
Resolution Spectrometer (GHRS) on board the Hubble Space Telescope (HST) 
show hot neutral H absorption along the line-of-sight
to a few nearby stars, including Sirius A and $\epsilon$ CMa (Bertin et al.\
1995, hereafter BVL95; Gry et al.\ 1995). The origin of this absorption,
however, has been a matter of debate. For the Sirius line-of-sight, BVL95
proposed that a hot conductive interface is responsible for excess Ly-$\alpha$
seen on the red side of the absorption profile, and Bertin et al.\ (1995b)
proposed that absorption from Sirius's wind accounted for excess Ly-$\alpha$
absorption on the blue side.  In other cases, detected hot H components are
undoubtedly linked to the neutral gas formed either around our own
heliosphere, due to the interaction of the solar wind with the ambient
interstellar medium of the Local Interstellar Cloud (LIC) (Linsky \& Wood
1996, hereafter LW96), or to neutral gas around other \textit{%
astrospheres} due to the corresponding interaction between the stellar winds
and the ambient neutral interstellar gas (Wood et al.\
1996). But for Sirius A and $\epsilon$ CMa, the combination of conductive
interface and stellar wind absorption has
remained the most likely source. 

There are 3 different types of heliospheric H atoms, in
addition to the unperturbed interstellar neutral H called primary 
interstellar atoms or PIA's: i) the compressed, decelerated, and heated
interstellar atoms (HIA's) formed by charge exchange with heated
interstellar protons outside the heliopause, ii) the neutralized, 
decelerated, and heated solar wind atoms (HSWA's) formed in the heliosheath
by charge exchange between the neutral interstellar gas and the hot
protons of the decelerated and compressed solar wind, and iii) the
neutralized supersonic solar wind atoms (SSWA's).  Only the HIA's and 
HSWA's are of interest here since the SSWA component is
flowing radially at very large velocities and will not produce absorption
in the central part of the Ly-$\alpha$ lines, and the PIA's are
indistinguishable from the normal interstellar gas.  The HIA's on the
upwind side of the heliosphere (i.e.\ the direction from which the
interstellar wind flows) collectively make up the so-called ``H-wall'',
which is the gas that has been detected towards $\alpha$ Cen (LW96).

The properties of the HSWA's are the most difficult to calculate,
since this hot gas has a very large mean free path and its characteristics
at one location in the heliosphere depend on the properties of all the
source regions everywhere in the heliosheath. Indeed, significant
differences between multi-fluid models and kinetic models have been found by
Williams et al.\ (1997). These authors have also
suggested that the mixing between the hot and warm populations
in the heliospheric tail through H-H collisions could be the origin of the
hot gas absorption observed towards Sirius. While recent computations show
that H-H collisions are negligible compared to charge-exchange processes
(Izmodenov et al.\ 1999b), our conclusions below will ultimately be similar
to their original idea.

The goal of this letter is to show that when one uses updated
parameters of the circumsolar interstellar medium and a very precise
kinetic/gasdynamic self-consistent model of the heliosphere, HSWA's produce
a non-negligible absorption in almost all directions, with a maximum effect
on the downwind side. We reconsider the Sirius A HST Ly-$\alpha$
spectrum and show that the red wing of the absorption is very well
fitted using our model. Then we show using simple analogies
that the additional absorption on the blue wing could be produced by HSWA's
and HIA's around Sirius itself, if the star is embedded in the neighboring
cloud detected towards the star by Lallement et al.\ (1994) and if the star
produces a wind, which is likely. 

\subsection{Heliospheric absorption towards Sirius}

A description of our self-consistent heliospheric model of the solar
wind-interstellar gas interaction can be found in Baranov \& Malama (1993),
Baranov et al.\ (1998), and Izmodenov et al.\ (1999a). We have updated the
interstellar parameters to take into account recent advances in the field,
such as the velocity and temperature determinations of the LIC from {\it in
situ} helium measurements and stellar spectroscopy (Witte et al.\ 1993;
Lallement \& Bertin 1992; Bertin et al.\ 1993), as well as estimates of the
neutral H and electron density in the circumsolar interstellar medium
(Lallement et al.\ 1996; Izmodenov et al.\ 1999a). In what follows, the 
assumed interstellar parameters are then: $T=6000$ K, $V=25$ km/s,
$N(HI)=0.2$ cm$^{-3}$, $N(e^{-})=0.07$ cm$^{-3}$. The upwind direction is
taken as $\lambda = 254.5^{\circ}$, $\beta = 7.5^{\circ}$ (ecliptic
coordinates), which translates into $ l_{II}= 186^{\circ }$,$ b_{II}=
-16^{\circ }$ (galactic coordinates). The assumed solar wind parameters at
1 AU are: $n(p)=7$ cm$^{-3}$, $V=450$ km/s. The model does not include
an interstellar magnetic field, but our estimates should not be
significantly changed in the presence of a moderate field. The boundary of
the model grid is at a distance of about 2000 AU in the direction of Sirius.

Fig.\ 1a is a sketch of the heliosphere and shows the direction of Sirius on
the downwind side. The predicted absorption by HSWA's and HIA's in the
direction of Sirius at an angle of $139^{\circ}$ from the
upwind direction is displayed in Fig.\ 1b. The absorption is shown in a
heliocentric rest frame.  It can be seen that the HSWA's are  the
main absorbers, and that their absorption is far from negligible. 

\begin{figure*}[tbp]
%\label{FigGam}\rule{0.4pt}{4cm}% line thickness, height of picture
%\psfig{figure=fig.epsf,height=12.45cm,width=18.0cm,angle=0} 
\psfig{figure=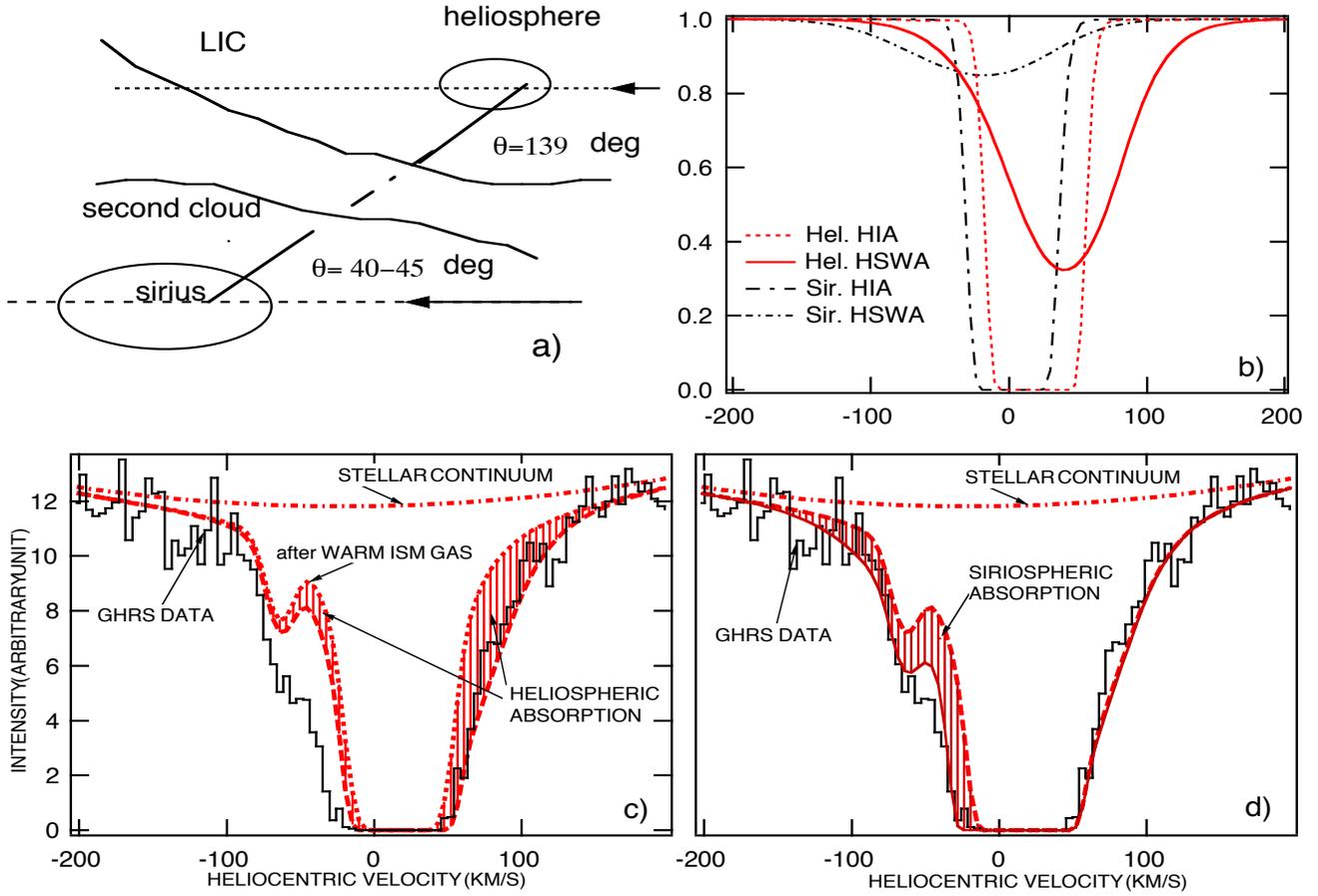,height=12.0cm,width=17.6cm,angle=0}
\caption{a) Schematic view of the Sun-Sirius line-of-sight. b) Transmission
as a function of Doppler shift in the solar rest frame through heliospheric
HIA's and HSWA's in the direction of Sirius, and through the siriospheric H
atoms in conditions described in the text. c) The GHRS spectrum, the
simulated profile after ISM absorption, and the profile after
ISM + heliospheric absorption. d) Simulated profile after
ISM + Heliospheric + Siriospheric absorption, where a good fit to the data
is obtained for a column density of HIA and HSWA atoms two times the heliospheric
column for the same orientation.}
%\hfill      \parbox[b]{5cm}{\caption[]{
%Neutral oxygen density in the upwind direction for
%different populations of interstellar neutrals.
%              }}%
\end{figure*}

\subsection{Heliospheric and interstellar absorption towards Sirius}

The 2.7 pc long line-of-sight to Sirius has been shown to cross two clouds:
i) the LIC, which in this direction is seen at a positive redshift of 19
km s$^{-1}$, and ii) a second cloud at a Doppler shift of 13 km s$^{-1}$
(Lallement et al.\ 1994), which is probably of the same type as our Local
Cloudlet.  Using the angularly close star $\epsilon$ CMa, Gry \& Dupin
(1998) have argued that the LIC extent in that direction is not longer
than $\thickapprox $ 0.6 pc.

Figure 1c shows the Sirius spectrum around the Ly-$\alpha $ line, and a
simple polynomial fit to the continuum surrounding the D and H absorption.
Superimposed on the data is the expected profile after absorption
by the two clouds at $V=13$ and 19 km s$^{-1}$, respectively, with an assumed
temperature of $T = 6000$ K.  The column densities
of the two clouds are both $1.6\times 10^{17}$ cm$^{-2}$, in agreement with the D absorption and a D/H ratio of $1.65\times 10^{-5}$ (Linsky et al, 1995,
BVL95).
Our conclusions are not sensitive
to either the exact value of D/H, or to the exact interstellar
temperature. The absorption has been convolved by the
instrumental profile corresponding to the G160M/SSA settings of the GHRS
spectrograph.

It is clearly seen that with warm interstellar gas only, absorption is
missing on both sides of the line, as already noticed by BVL95.  After
adding the modeled absorption by the heliosphere, the resulting profile is
substantially modified on the red part of the line. To make the heliospheric
effect clear in Fig.\ 1c, the additional absorption is shown as a hatched
area. It can be seen that the red part of the observed spectrum is well
fitted by the model.  Thus, while the heliosphere cannot been made
responsible for the absorption in the blue wing, there is {\em no need to
propose additional absorption from interstellar hot gas} along the
line-of-sight to fit the red side of the absorption line.

\section{Heliospheric, interstellar, and \textit{Siriospheric }absorption
towards Sirius}

Bertin et al.\ (1995b) have suggested that the additional absorption on the
blue side is due to neutral gas associated with Sirius's wind, a counterpart
of Mg~II absorption detected at this velocity.  Here we consider another
possibility, that the absorption is from the interaction area between
Sirius's wind and the interstellar gas around the star.  

In the following, we make a series of assumptions:

- Sirius is embedded in the ``blue cloud'' seen at the Doppler shift of 13
km s$^{-1}$. This is a very reasonable assumption, owing
to the very small length of the line-of-sight.

- Sirius has a wind with a terminal velocity of the same order of magnitude
as the solar wind velocity (say 400--1500 km s$^{-1}$), and a mass flux
at least that of the solar wind. These assumptions are compatible with the
predictions of radiatively driven wind models or coronal winds
(see Bertin et al. 1995b).

- The gas near the Siriopause is not fully ionized by the EUV radiation from
Sirius B.  Using model results of Paerels et al. (1987) for a 25,000 K pure
H white dwarf, the
EUV flux of Sirius B balances the travel time associated to a star/ISM
relative motion of $\thickapprox $25 km s$^{-1}$ at distances of about 200
AU, which implies that if the size of the siriosphere is of this order or
larger, neutral atoms of the cloud can penetrate within it. Such a size is
very likely reached, since the Sirius wind is probably stronger than
the solar wind and then the equilibrium with the ISM is reached at larger
distances.

- The axis of symmetry of the siriosphere, determined by the relative motion
between the star and the ambient gas, makes an angle
$\theta \thickapprox 40^{\circ}$ with the line-of-sight direction. 
We know the 3D motion of Sirius A from ephemerides for the
orbital system, combined with the radial velocity of Sirius A at the time of
the observations, v$_{r}$= -5 km s$^{-1}$ , but we do
not know the 3D motion of the cloud.
Multiple clouds have been observed for many short lines of sight besides
Sirius, but their projected velocities are never far from that of the LIC;
the separation is only 6 km s$^{-1}$ for the non-LIC cloud seen towards
Sirius.  Thus, assuming the motions of these additional clouds are identical
to that of the LIC is a reasonable approximation, and making this assumption
for the Sirius cloud leads to an estimated
angle of $\theta\approx 40^{\circ}$.

Under these assumptions, we can estimate some characteristics of the HIA and
HSWA populations around Sirius. The distance at which pressure equilibrium
between the wind and the ISM is reached depends on the ISM pressure and the
stellar wind momentum flux. If the mass flux and/or the velocity
of the Sirius wind are larger than the solar wind flux and velocity,
which is likely, the HIA component will be created at larger distances from
the star in comparison with the solar case. But for the interstellar gas
outside the discontinuity, the conditions of deceleration and heating should
be about the same as for the heliosphere, since the gas has to decelerate in
both cases by about the same quantity to be at rest with the star (the
relative velocity between the star and the ISM). In the solar case, the
relative velocity is 25.5 km s$^{-1}$.  For Sirius, we can estimate this
velocity to be about 20--40 km s$^{-1}$.  The velocity is $\approx 31$
km s$^{-1}$ if the surrounding cloud motion is assumed to be identical to
the LIC.  However, since it's projected velocity (13 km s$^{-1}$) is lower
than the LIC's projected velocity (19 km s$^{-1}$), the actual relative
velocity is likely to be lower than 31 km s$^{-1}$ and therefore not too
different from the solar case.  In this instance, the absorption will be
found at velocities between 13 km s$^{-1}$ (projection of the cloud's
velocity) and $-5$ km s$^{-1}$ (the projection of the stellar motion).
Fig.\ 1b shows the resulting theoretical absorption. 

The compressed stellar wind should also have properties similar to the
compressed solar wind, although possibly formed at larger distances and
possibly hotter. We have computed the absorption in the solar frame for
$\theta = 45^{\circ}$, which should be equivalent to what would be seen by
an observer on Sirius. Then, we have changed its sign and added $-5$
km s$^{-1}$ to represent what would be seen for an observer at
rest with the Sun and looking towards Sirius. Fig.\ 1b shows the predicted
absorption. It can be seen that the HIA and especially the
HSWA absorptions fall at the location of the ``missing'' absorption in the
blue wing. 

Figure 1d shows the consequences of this additional absorption on the
simulated spectrum. In order to obtain a complete ``filling'' of the line we
have multiplied by 2 the column density of the HIA and HSWA components, which
corresponds to a cloud two times denser than the LIC, or distances
in the siriosphere two times larger, or any combination. It is beyond
the scope of this paper to investigate all solutions since there are too
many.  However, from this crude estimate we conclude that
\textit{siriospheric }absorption could possibly account for the extra
absorption observed in the blue wing.

\section{Conclusion and discussion}

We have used the most precise and updated model of the Sun-LIC interaction
to calculate the Ly-$\alpha$ absorption by the neutral gas in and around
the heliosphere along the line-of-sight towards Sirius. We find that the
neutralized solar wind from the heliosheath is mainly responsible for the
absorption, and that the red side of the absorption line is very well fitted
when adding this absorption to the normal interstellar absorption. In these
conditions, there is no need to propose
interstellar hot gas from a conductive interface to explain the red wing
absorption, as BVL95 did in their analysis.

Using analogies with the solar case, we also show that the remaining
missing absorption on the blue side could be explained in the same way by a
``siriosphere'', if Sirius is embedded in the neighboring cloud seen towards
the star, if it has a stronger wind than the Sun, and if Sirius B does not
completely ionize the hydrogen in and around the siriosphere.
In this interpretation, there is no need for neutral H associated with a
supersonic wind like that proposed by Bertin et al.\ (1995b).

We also point out that the model results show that heliospheric absorption
cannot be neglected in {\em any} Ly-$\alpha$ analysis, whatever the
line-of-sight direction, if the interstellar absorption is relatively low
(N(HI) $\leq 10^{18.5}$ cm$^{-2}$). 

\begin{acknowledgements}

We thank our referee Brian Wood for his valuable comments, his help 
for the rewording of the paper and the improvements he suggested.
This work has been done in the frame of the INTAS project 
"The heliosphere in the Local Interstellar Cloud" 
and partly supported by the ISSI in Bern. 
V.I. has been supported by the MENESR (France).
Yu.M. has been supported by RFBR Grants No.98-01-00955,98-02-16759.
\end{acknowledgements}

%--------------------------------FIGURES--------------------------------
%
%\includegraphics{JGR_960425.ps}
%\special{psfile=JGR_970302.ps hoffset=-100 voffset=100 hscale=100 vscale=100}
%\begin{figure} 
%\psfig{figure=JGR_960425.ps,height=10.5cm,width=15.0cm}
%\psfig{figure=chiv2.ps,height=10.5cm,width=15.0cm}

%\caption{Representation of the 
%shown by crosses.}
%\end{figure}

\end{document}